\documentclass[10pt,preprint,nocopyrightspace]{sigplanconf}

\usepackage{graphicx}
\usepackage{appendix}
\usepackage{colonequals}
\usepackage{stmaryrd}
\usepackage{latexsym}
\usepackage{mathpartir}

\usepackage{xspace}
\usepackage{amsmath}
\usepackage{defs}
\usepackage{color}
\usepackage{url}

\usepackage{algorithm}
\usepackage[noend]{algpseudocode}

\algnotext{EndFor} 
\algnotext{EndIf}
\algrenewcommand\algorithmicindent{0.8em}%

\usepackage{listings}

\lstdefinelanguage{ML}{
  alsoletter={*},
  morekeywords={datatype, of, if, *},
  sensitive=true,
  morecomment=[s]{/*}{*/},
  morestring=[b]"
}

\lstdefinelanguage{scala}{
  alsoletter={@,=,>},
  morekeywords={abstract, case, catch, choose, class, def, do, else, extends, false, final, finally, for, if, implicit, import, match, new, null, object, let,
override, package, private, protected, requires, return, sealed, super, this, throw, trait, try, true, type, val, var, while, with, yield, domain, 
postcondition, precondition,invariant, constraint, assert, forAll, _, return, @generator, ensure, require, ensuring, assuming, otherwise, asserting}
  sensitive=true,
  morecomment=[l]{//},
  morecomment=[s]{/*}{*/},
  morestring=[b]"
}

\newcommand{\codestyle}{\small\sffamily}

\newcommand{\SAND}{\mbox{\tt \&\&}\xspace}

\newcommand{\PP}{\mbox{\tt ++}\xspace}
\newcommand{\MM}{\mbox{\tt {-}{-}}\xspace}
\newcommand{\RA}{\Rightarrow}
\newcommand{\EQ}{\mbox{\tt ==}}

\newcommand{\SLE}{\ensuremath{\leq}}
\newcommand{\SGE}{\ensuremath{\geq}}

\begin{document}


\newcommand{\ourtitle}{On Integrating Deductive Synthesis and Verification Systems}
\titlebanner{Technical Report, April 2013}      
\preprintfooter{\ourtitle}   

\title{\ourtitle}

\authorinfo{Etienne Kneuss \and Viktor Kuncak \and Ivan Kuraj \and Philippe Suter}
           {{\'E}cole Polytechnique F{\'e}d{\'e}rale de Lausanne (EPFL), Switzerland}
           {\texttt{firstname.lastname@epfl.ch}}

\maketitle
\sloppy

\begin{abstract}
We describe techniques for synthesis and verification of
recursive functional programs over unbounded domains. Our
techniques build on top of an algorithm for satisfiability
modulo recursive functions, a framework for deductive
synthesis, and complete synthesis procedures for algebraic
data types. We present new counterexample-guided algorithms
for constructing verified programs. We have implemented
these algorithms in an integrated environment for
interactive verification and synthesis from relational
specifications. Our system was able to synthesize a number
of useful recursive functions that manipulate unbounded
numbers and data structures.

\end{abstract}


\section{Introduction}

Software construction is a difficult problem-solving
activity. It remains a largely manual effort today, despite
significant progress in software development environments
and tools. 
The development becomes even more difficult 
when the goal is to deliver \emph{verified} software,
which must satisfy specifications such as
assertions, pre-conditions, and post-conditions. 

We believe that quick feedback and error reports are essential
for practical verification.
Verifying programs after they have been developed
is extremely time-consuming \cite{DBLP:conf/fm/LeinenbachS09,
KleinETAL09seL4FormalVerificationOSKernel} and it is difficult to argue
its cost-effectiveness.
Our research therefore explores approaches that support \emph{integrated software construction
and verification}. An important aspect of such approaches
are modular verification techniques which can check that a function
conforms to its local specification. 
In such approach,
the verification of an individual function against its specification can start before the entire
software system is completed, so
tools can provide rapid feedback that allows specifications
and implementations to be developed simultaneously. 
Quoting 
\cite{Barnett:2011:SVS:1953122.1953145}, who report on the experience with Spec\#,
\begin{quote}
``If verification ever makes it into the daily rhythm of mainstream programming, it will be through a design-time interface providing online verification.''
\end{quote}

We choose a functional language
as the core language for the development of verified software.
Functional languages strike
an appealing balance between executability and verifiability, predicted already in \cite{McCarthy60Recursive}.
Although the problem of delivering verified software has
been explored through a number of different approaches, in a number of successful cases
the development relies heavily on a functional language.
In some cases \cite{KleinETAL09seL4FormalVerificationOSKernel} the researchers
have even written the entire software system once in a functional language for verifiability, and once in a
lower-level language for execution efficiency.

Based on the ideas of suitability of a functional paradigm and
the importance of rapid feedback, we have developed a
verifier that quickly detects errors in functional programs
and reports concrete counterexamples, but can also prove the
correctness of programs
\cite{Suter12ProgrammingSpecifications,
  SuterDottaKuncak10DecisionProceduresAlgebraicDataTypesAbstractions,
  SuterKoeksalKuncak11SatisfiabilityModuloRecursivePrograms}.
Furthermore, we have integrated such countexample-generating
verifier into a web-browser-based IDE, resulting in a tool for
convenient development of verified functional programs.
This verifier is the starting point of the tool we present in this
paper. 

Moving beyond verification, we believe that a productive development
of verified software requires 
techniques for synthesis from specifications.
Specifications in terms of properties generalize existing declarative programming language
paradigms by allowing the statement of \emph{constraints} between inputs and outputs
as opposed to always specifying outputs as functions of inputs 
\cite{JaffarLassez87ConstraintLogicProgramming,
  KoeksalKuncakSuter12ConstraintsAsControl}.
Unlike deterministic specifications, constraints can be composed using conjunctions, which
enables description of the problem as a combination of orthogonal requirements.

This paper introduces synthesis algorithms, techniques and tools
that integrate synthesis into the development process for functional program.
We present a synthesizer that can construct the bodies of functions starting
solely from their postconditions. 
The programs that our synthesizer produces typically
manipulate unbounded data types, such as algebraic data
types and unbounded integers.  Thanks to the use of deductive
synthesis and the availability of a verifier, when
the synthesizer succeeds, the generated code is guaranteed to
be correct for all possible input values. 

Our synthesizer uses specifications as the description of
the synthesis problems. While it could additionally accept
input/output examples to illustrate the desired
functionality, we view such illustration as a special form
of input/output relation: whereas input/output examples
correspond to tests and provide a description of a finite
portion of the desired functionality, we primarily focus on
symbolic descriptions, which ensure the desired behavior
over an arbitrarily large or even infinite domain. From such
descriptions, our synthesizer can automatically generate
input/output examples when needed, but can also use them and
transform them directly into executable code. 

A notable degree of automation in our
synthesizer comes from synthesis procedures
\cite{KuncakETAL10CompleteFunctionalSynthesis,KuncakETAL12SoftwareSynthesisProcedures,
JacobsKuncakSuter13ReductionsSynthesisProcedures}, which
compile specification fragments expressed 
in decidable logics. Our work is the first implementation of the synthesis procedure
for algebraic data types \cite{Suter12ProgrammingSpecifications}.

Note however, that, to capture a variety of scenarios in software development,
we also support the general problem of synthesis
of Turing-complete programs. 
The result is a framework for cost-guided
application of deductive synthesis rules, which decompose the problems into
subproblems.

Our synthesizer tightly cooperates with the underlying
verifier, which allows it to achieve orders of magnitude
better performance than using simpler generate-and-test
approaches. Techniques we use include symbolic transformation
based on synthesis procedures,
as well as synthesis of recursive functions using counterexample-guided strategies.
We have evaluated a number of system architectures and trade-offs between
symbolic and concrete reasoning in our implementation and arrived at an implementation
that is successful despite the large space of possible programs.


We believe that we have
achieved a new level of automation for a broad
domain of recursive functional programs. We consider a particular 
strength of our system that it can synthesize code that
satisfies a given relational specification for all values of inputs, and not
only given input/output pairs.

Despite aiming at a high automation level, we are aware that any general-purpose automated synthesis
procedure will ultimately face limitations of scalability and the ability to control
the development process.  
We deployed the synthesis algorithm as an interactive assistance
that allows the developer to interleave manual and automated development
steps. In our system,
the developer can decompose a function and leave the subcomponents
to the synthesizer, or, conversely, the synthesizer can decompose the problem,
solve some of the subproblems, and leave the remaining open cases for the developer.
To facilitate such synergy, we deploy an anytime synthesis
procedure, which maintains a ranked list of current problem
decompositions. The user can interrupt the synthesizer at any time to display
the current solution and continue manual development. This is possible thanks
to the fact that synthesis problems and specification problems are both
expressed in a unified language based on the construct resembling non-deterministic choice.

\subsection{Contributions}

The overall contribution of this paper is an integrated synthesis and development
system for automated and interactive development of verified programs. A number
of techniques from deductive and inductive reasoning need to come together to make
such system usable. 

\paragraph{Verifier.} 
Our automated verification environment is the enabler
  of synthesis.  We use SMT solvers, specifically Z3
  \cite{MouraBjoerner08Z3}, and a fair function unfolding
  strategy that is effective for sufficiently surjective
  abstractions
  \cite{SuterDottaKuncak10DecisionProceduresAlgebraicDataTypesAbstractions,
    SuterKoeksalKuncak11SatisfiabilityModuloRecursivePrograms}.
  We have achieved substantial speed-ups of this technique
  for satisfiable constraints through the use of code
  generation and fair enumeration of structured values. The improvements
  in verification and falsification have transferred to the improvements
  in synthesis times.

\paragraph{Implemented synthesis framework.}
We developed a deductive synthesis framework that can accept a given set of
synthesis rules and applies them according to a cost function. The framework
accepts 1) a path conditions that encode program context, and
2) a relational specifications. It returns the function from inputs to outputs as a solution,
as well as any necessary strengthening
of the precondition needed for the function to satisfy the specification. We have deployed 
the framework in a web-browser-based environment with
continuous compilation and the ability to interrupt the synthesis to obtain a partial
solution in the form of a new program with a possibly simpler synthesis problem.

\paragraph{Data type synthesis.}
Within the above framework we have implemented
rules for synthesis of algebraic data type equations and disequations 
\cite{Suter12ProgrammingSpecifications}, as well
as a number of general rules for decomposing specifications based on their
logical structure or case splits on commonly useful conditions.
We have developed program simplification techniques that post-process
the generated code and make it more readable.

\paragraph{Support for recursion schemas and symbolic term generators.} 
One of the main strengths in our framework is a new form of
counterexample-guided synthesis that arises from a combination of several
rules.
\begin{itemize}
\item A set of built-in recursion schemas can solve a problem
by generating a fresh recursive function. To ensure well-foundedness we have extended
our verifier with termination checking, and therefore generate only terminating function
calls in this rule.
\item To generate bodies of functions, we have symbolic term generators that systematically
generate well-typed programs built from selected set of operators (such as algebraic data type
constructors and selectors).
To test candidate terms against specifications we use the {\leon}'s verifier. To speed up this search,
the rule accumulates previously found counterexamples. Moreover, to quickly bootstrap the set of
examples it uses systematic generators that can enumerate in a fair way any finite prefix
of a countable set of structured values. The falsification of generated bodies is done by direct
execution of code. For this purpose, we have developed a lightweight compiler for our subset of Scala
into bytecodes, replacing many constraint reasoning steps by code execution.
\end{itemize}

\paragraph{Function generation by condition abduction.}
We also present and evaluate an alternative counterexample-guided rule tailored
towards synthesis of recursive conditional functions, with the following characteristics.
\begin{itemize}
\item Instead of specialized term evaluators, the rule uses a general expression enumerator
based on generating all expressions of a given type \cite{Gvero13CompleteCompletionUsingTypesAndWeights}.
This results in a broad coverage of expressions
that the rule can synthesize. It uses a new lazy enumeration algorithm for such
expressions with polynomial-time access to the next term to enumerate 
\cite{Kuraj13InteractiveCodeGeneration}.
Similarly to the previous rule, it filters well-typed expressions using counterexamples generated
from specifications and previous function candidates, as well as based on structured value genererators.
\item The most distinctive aspect of this rule is the handling of conditional expressions. The expressions
are synthesized by collecting relevant terms that satisfy a notable number of derived test inputs, and
then trying to synthesize predicates that imply the correctness of candidate terms. This is an alternative
to relying on existing rules to perform splitting on simple conditions.
Effectively, the additional rule performs abduction of conditions until it covers the entire 
input space with a partition of conditions, where each partition is
associated with a term.
\end{itemize}

\paragraph{Evaluation.}
We evaluate the current reach of our synthesizer in fully automated mode by synthesizing
functions such as those
that merge, partition, and sort lists of objects, where lists are defined using a general
mechanism for algebraic data types.
This paper presents a description of all the above techniques and a snapshot
of our results. We believe that the individual techniques are interesting by themselves, but 
we also believe that having a system that combines them is essential to understand
the potential of these techniques in addressing the difficult problem as synthesis.
To gain full experience of the feeling of such development process,
we therefore invite the reader to explore the system themselves.



\section{Interactive Synthesis and Verification in the Leon System}
\label{sec:examples}

We start by illustrating through a series of examples how developers can
leverage our system to write programs that are correct by construction.

\paragraph{Unary numerals.}
As a first example, we will consider tasks related to Unary numerals. While
these examples are simple in nature, they illustrate some very important
points. In particular, they show how, using a combination of verification and
synthesis, one can program functions manipulating data types in one
representation while specifying the operations using an abstract view.

Consider a standard definition of unary numerals as a recursive data type,
with a base case ``zero'' and a ``successor'' constructor.
\begin{lstlisting}
sealed abstract class Num
case object Z extends Num
case class  S(pred: Num) extends Num
\end{lstlisting}
Because it is more convenient to think of these numerals in term of their
integer value, we can define an abstraction function that computes it:
\begin{lstlisting}
def value(n:Num) : Int = (n match {
  case Z $\RA$ 0
  case S(p) $\RA$ 1 + value(p)
}) ensuring (_ $\ge$ 0)
\end{lstlisting}

The \cl{ensuring} clause is Scala notation for a 
postcondition~\cite{WorkaroundOdersky10ContractsScala}. 
These postconditions are defined by an
anonymous function, whose single argument denotes the result of the function.
The underscore notation is a shorthand for \cl{x $\RA$ x $\ge$ 0}, so this
annotation simply specifies that the integer representation of a unary numeral
is never negative, and \leon instantly proves this simple verification condition.

Using our (verified) abstraction function, we can start specifying operations
on unary numerals. Consider for instance the addition operation. Its contract
in terms of the \cl{value} function is clear, so we can write it as:
\begin{lstlisting}
def add(x : Num, y : Num) : Num = choose { (r: Num) $\RA$
  value(r) $\EQ$ value(x) + value(y)
}
\end{lstlisting}
Here, \cl{choose} is a special function defined by \leon to represent a
computation that needs to be synthesized. Similarly to the postcondition, it is
defined by an anonymous functions whose result represents the desired output.
Contrary to postconditions, though, the function (or \emph{synthesis
predicate}) can admit multiple arguments, in which case the synthesized program
should return a tuple of values of the appropriate types.

Upon invocation of the \leon synthesis component, the following recursive
implementation is derived:
\begin{lstlisting}
def add(x : Num, y : Num) : Num = (x match {
  case Z $\RA$ y
  case S(p) $\RA$ add(p, S(y))
}) ensuring(r $\RA$ value(r) $\EQ$ value(x) + value(y))
\end{lstlisting}
We can continue expanding on these results, and define a synthesis predicate
for multiplication:
\begin{lstlisting}
def mult(x : Num, y : Num) : Num = choose { (r: Num) $\RA$
  value(r) $\EQ$ value(x) * value(y)
}
\end{lstlisting}
Leveraging the previous results for \cl{add}, our synthesis algorithm derives
the following program:
\begin{lstlisting}
def mult(x : Num, y : Num) : Num = (x match {
  case Z $\RA$ Z
  case S(p) $\RA$ add(y, mult(p, y))
}) ensuring(r $\RA$ value(r) $\EQ$ value(x) * value(y))
\end{lstlisting}
Both functions are generated within three seconds.

\paragraph{List manipulation.}
We believe this rapid feedback is mandatory when developing from
specifications. One reason is that, since contracts are typically partial,
results obtained from under-specifications can be remote from the desired
output. Thus, a desirable strategy is to rapidly iterate and refine
specifications until the output matches the expectations.

As an example, we will consider the task of synthesizing the \cl{split}
function necessary in merge sort. We start from a standard recursive definition
of lists, and we assume the existence of recursive functions computing their
size (as a non-negative integer), and their content (as a set of integers).
\begin{lstlisting}
sealed abstract class List
case class Cons(head: Int, tail: List) extends List
case object Nil extends List

def size(lst : List) : Int = $\ldots$
def content(lst : List) : Set[Int] = $\ldots$
\end{lstlisting}

As a first attempt to synthesize \cl{split}, we try the following specification:
\begin{lstlisting}
def split(lst : List) : (List,List) = choose { (r : (List,List)) $\RA$
  content(lst) $\EQ$ content(r._1) $\PP$ content(r._2)
}
\end{lstlisting}
\leon instantly generates the following function which, while it satisfies the contract,
is not particularly useful:
\begin{lstlisting}
def split(lst : List) : (List,List) = (lst, Nil)
\end{lstlisting}
To avoid getting a single list with an empty one, we can refine the
specification by enforcing that the sizes of the resulting lists should not
differ by more than one:
\begin{lstlisting}
def split(lst : List) : (List,List) = choose { (r : (List,List)) $\RA$
  content(lst) $\EQ$ content(r._1) $\PP$ content(r._2)
  $\SAND$ abs(size(r._1)-size(r._2)) $\SLE$ 1
}
\end{lstlisting}
Again, \leon instantly generates a correct, useless, program:
\begin{lstlisting}
def split(lst : List) : (List,List) = (lst, lst)
\end{lstlisting}
We can further refine the specification by stating that the \emph{sum} of the
sizes of the two lists should match the size of the input one:
\begin{lstlisting}
def split(lst : List) : (List,List) = choose { (r : (List,List)) $\RA$
  content(lst) $\EQ$ content(r._1) $\PP$ content(r._2)
  $\SAND$ abs(size(r._1) - size(r._2)) $\SLE$ 1
  $\SAND$ (size(r._1) + size(r._2)) $\EQ$ size(lst)
}
\end{lstlisting}
We then finally obtain a useful \cl{split} function:
\begin{lstlisting}
def split(lst: List): (List, List) = lst match {
  case Nil $\RA$ (Nil, Nil)
  case Cons(h, Nil) $\RA$ (Nil, Cons(h, Nil))
  case Cons(h1, Cons(h2, t2)) $\RA$
    val r = split(t2)
    (Cons(h1, r._1), Cons(h2, r._2))
}
\end{lstlisting}
We observe that in this programming style, users can write (or generate) code
by conjoining orthogonal requirements, such as constraints on the sizes and
contents, which are only indirectly related. The rapid feedback make it
possible to go through multiple candidates rapidly, strengthening the
specification as required.

\paragraph{Sorting.}
A typical example of a task that is easier to specify than to implement is
sorting. We conclude this overview of \leon's synthesis capabilities by showing
how to derive an insertion sorting algorithm. We start from the straightforward
definition of \cl{isSorted}, a function that \emph{checks} whether a list is
sorted:
\begin{lstlisting}
def isSorted(lst : lst) : Boolean = lst match {
  case Nil $\RA$ true
  case Cons(_, Nil) $\RA$ true
  case Cons(x1, xs @ Cons(x2, _)) $\RA$ x1 $\le$ x2 $\SAND$ isSorted(xs)
}
\end{lstlisting}
Using this function, the problem of sorting can be stated as simply as:
\begin{lstlisting}
def sort(lst : List): List = choose { (r: List) $\RA$
  isSorted(r) $\SAND$ content(r) $\EQ$ content(lst)
}
\end{lstlisting}
To achieve this goal, we start by specifying the helper function \cl{insertSorted}:
\begin{lstlisting}
def insertSorted(lst : List, v: Int): List = {
  require(isSorted(lst))
  choose { (r: List) $\RA$
    isSorted(r) $\SAND$ content(r) $\EQ$ content(lst) $\PP$ Set(v)
  }
}
\end{lstlisting}
From this, \leon generates the following solution:
\begin{lstlisting}
def insertSorted(lst: List, v: Int): List = {
  require(isSorted(lst))
  lst match {
    case Nil $\RA$ Cons(v, Nil)
    case Cons(h, tail) $\RA$
      val r = insertSorted(t, v)
      if (v > h) Cons(h, r)
      else if (h $\EQ$ v)  r
      else Cons(v, Cons(h, t))
}
\end{lstlisting}
With the help of this insertion function, we can proceed to synthesizing
\cl{sort} with the simple specification mentioned above.  Within five seconds,
\leon generates the following implementation of insertion sort:
\begin{lstlisting}
def sort(lst : List): List = lst match {
  case Nil $\RA$ Nil
  case Cons(h, t) $\RA$ insertSorted(sort(t), h)
}
\end{lstlisting}

\section{The Leon Verifier}
\label{sec:leonverifier}

The results presented in this paper focus on the synthesis component of \leon.
%
The language of \leon is a subset of Scala, as illustrated through the examples
of Section~\ref{sec:examples}. Besides integers and user-defined recursive data
types, \leon supports booleans, sets and maps.

\paragraph{Solver algorithm.} At the core of \leon is an algorithm to reason
about formulas that include user-defined recursive functions, such as
\cl{size}, \cl{content}, and \cl{isSorted} in Section~\ref{sec:examples}. The
algorithm proceeds by iteratively examining longer and longer execution traces
through the recursive functions. It alternates between an over-approximation of
the executions, where only unsatisfiability results can be trusted, and an
under-approximation, where only satisfiability results can be concluded. The
status of each approximation is checked using the state-of-the-art SMT solver
Z3 from Microsoft Research \cite{MouraBjoerner08Z3}. The algorithm is a
\emph{semi-decision procedure}, meaning that it is theoretically complete for
counterexamples: if a formula is satisfiable, \leon will eventually produce a
model \cite{SuterKoeksalKuncak11SatisfiabilityModuloRecursivePrograms}.
Additionally, the algorithm works as a decision procedure for a certain class
of formulas
\cite{SuterDottaKuncak10DecisionProceduresAlgebraicDataTypesAbstractions}.

In the past, we have used this core algorithm in the context of verification
\cite{SuterKoeksalKuncak11SatisfiabilityModuloRecursivePrograms}, but also as
part of an experiment in providing run-time support for declarative programming
using constructs similar to \cl{choose}
\cite{KoeksalKuncakSuter12ConstraintsAsControl}. We have in both cases found
the performance in finding models to be suitable for the task at
hand. \footnote{We should also note that since the publication of
\cite{SuterKoeksalKuncak11SatisfiabilityModuloRecursivePrograms}, our
engineering efforts as well as the progress on Z3 have improved running
times by 40\%.}

Throughout this paper, we will assume the existence of an algorithm for deciding
formulas containing arbitrary recursive functions. Whenever completeness is an
issue, we will mention it and describe the steps to be taken in case of, e.g.
timeout.

\paragraph{Compilation-based evaluator.} Another component of \leon on which we
rely in this paper is an interpreter based on on-the-fly compilation to the
JVM. Function definitions are typically compiled once and for all, and can
therefore be optimized by the JIT compiler. This component is used during the
search in the core algorithm, to validate models and to sometimes
optimistically obtain counterexamples. We use it to quickly
reject candidate programs during synthesis (see sections~\ref{sec:cegis} and \ref{sec:insynth}).

\paragraph{Ground term generator.} Our system also leverages \leon's generator
of ground terms and its associated model finder. Based on a generate-and-test
approach, it can generate small models for formulas by rapidly and fairly
enumerating values of any type. For instance, enumerating \cl{List}s will produce a stream of values
\cl{Nil()}, \cl{Cons(0, Nil())}, \cl{Cons(0, Cons(0, Nil()))}, \cl{Cons(1,
Nil())}, \ldots

\section{Deductive Synthesis Framework}
\label{sec:framework}

The approach to synthesis we follow in this paper is to derive programs by a
succession of independently validated steps. In this section, we briefly
describe the formal reasoning behind these constructive steps and provide some
illustrative examples. A more extended exposition of this formal framework is
available in \cite{JacobsKuncakSuter13ReductionsSynthesisProcedures}.

\subsection{Synthesis Problems}
A synthesis problem is given by a predicate describing a desired relation
between a set of input and a set of output variables, as well as the context
(program point) at which the synthesis problem appears. We represent such a
problem as a quadruple
\[
  \br{\seqa}{\pcname}{\phi}{\seqx}
\]
where:
\begin{itemize}
  \item $\seqa$ denotes the set of \emph{input variables},
  \item $\seqx$ denotes the set of \emph{output variables},
  \item $\phi$ is the \emph{synthesis predicate}, and
  \item $\pcname$ is the \emph{path condition} to the synthesis problem.
\end{itemize}
The free variables of $\phi$ must be a subset of $\seqa \cup \seqx$. The path
condition denotes a property that holds for input at the program point where
synthesis is to be performed, and the free variables of $\pcname$ should
therefore be a subset of $\seqa$.

As an example, consider the following call to \choosesym:
\begin{lstlisting}
def f(a : Int) : Int = {
  if(a $\SGE$ 0) {
    choose((x : Int) $\RA$ x $\SGE$ 0 $\SAND$ a + x $\SLE$ 5)
  } else $\ldots$
}
\end{lstlisting}
The representation of the corresponding synthesis problem is:
\begin{equation}\label{eq:synprobdemo}
  \br{a}{a \ge 0}{x \ge 0 \land a + x \le 5}{x}
\end{equation}

\subsection{Synthesis Solutions}
We represent a solution to a synthesis problem as a pair
\[
  \pg{\prename}{\pgname}
\]
where:
\begin{itemize}
  \item $\prename$ is the \emph{precondition}, and
  \item $\pgname$ is the \emph{program term}.
\end{itemize}
The free variables of both $\prename$ and $\pgname$ must range over $\seqa$.
The intuition is that, whenever the path condition and the precondition are
satisfied, evaluating $\phi[\seqx \mapsto \pgname]$ should evaluate to
{\btrue}, i.e.\ $\pgname$ are realizers for a solution to $\seqx$ in $\phi$
given the inputs $\seqa$. Furthermore, for a solution to be as general as
possible, the precondition must be as weak as possible.

Formally, for such a pair to be a solution to a synthesis problem, denoted as
\[
  \br{\seqa}{\pcname}{\phi}{\seqx} \vdash \pg{\prename}{\pgname}
\]
the following two properties must hold:
\begin{itemize}
  \item \emph{Relation refinement:}
  \[
    \pcname \land \prename \models \phi[\seqx\mapsto\pgname]
  \]
  This property states that whenever the path- and precondition hold, the
program $\pgname$ can be used to generate values for the output variables
$\seqx$ such that the predicate $\phi$ is satisfied.

  \item \emph{Domain preservation:}
  \[
    \pcname \land (\EX{\seqx}{\phi}) \models \prename
  \]
  This property states that the precondition $\prename$ cannot exclude inputs
  for which an output would exist such that $\phi$ is satisfied.
\end{itemize}

As an example, a valid solution to the synthesis problem \eqref{eq:synprobdemo}
is given by:
\[
  \pg{a \le 5}{0}
\]
The precondition $a \le 5$ characterizes exactly the input values for which a
solution exists, and for all such values, the constant $0$ is a valid solution
term for $x$. Note that the solution is in general not unique; alternative
solutions for this particular problem include for instance $\pg{a \le 5}{5 -
a}$, or $\pg{a \le 5}{\mcl{if(}a < 5\mcl{)~} a + 1 \mcl{~else~} 0}$.

\paragraph{A note on path conditions.} 
Strictly speaking, the inclusion of the
path condition does not add expressive power to the representation of
synthesis problems. One
can easily verify that the space of solution terms for
$\br{\seqa}{\pcname}{\phi}{\seqx}$ is isomorphic to the one for
$\br{\seqa}{\btrue}{\pcname \land \phi}{\seqx}$. In the latter case, the path
condition $\pcname$, is simply 
included in the precondition of the solution. 
On the other hand, from the definition it follows
that if $\pg{\prename}{\pgname}$ is a solution and $\pcname \land \prename$
is equivalent to $\pcname \land \prename'$ then 
$\pg{\prename'}{\pgname}$ is also a solution to the synthesis problem.
We can let, $\prename'$ be, for example, $\pcname \land \prename$, or, as another extreme,
$\pcname \rightarrow P$.
We have therefore found it convenient 
in the implementation to explicitly keep track of the path conditions and allow
freedom in the representation of the returned precondition $\prename$.

\subsection{Inference Rules for Synthesis}
\begin{figure*}
\begin{mathpar}
\inferrule*[Left=One-point]
  {
    \br{\seqa}{\pcname}{\phi[x_0 \mapsto t]}{\seqx} \vdash \pg{\prename}{\pgname}
    \and
    x_0 \notin \varsof{t}
  }
  {
    \br{\seqa}{\pcname}{x_0 = t \land \phi}{x_0 \splus \seqx} \vdash \pg{\prename}{\pglet{\seqx}{\pgname}{(t \splus \seqx)}}
  }
\and
\inferrule*[Left=Ground]
  {
    \mathcal{M} \models \phi \and \varsof{\phi} \cap \seqa = \emptyset
  }
  {
    \br{\seqa}{\pcname}{\phi}{\seqx} \vdash \pg{\btrue}{\mathcal{M}}
  }
\and
\inferrule*[Left=Case-split]
  {
    \br{\seqa}{\pcname}{\phi_1}{\seqx} \vdash \pg{\prename_1}{\pgname_1}
    \and
    \br{\seqa}{\pcname \land \neg \prename_1}{\phi_2}{\seqx} \vdash \pg{\prename_2}{\pgname_2}
  }
  {
    \br{\seqa}{\pcname}{\phi_1 \lor \phi_2}{\seqx} \vdash \pg{\prename_1 \lor \prename_2}{\pgite{\prename_1}{\pgname_1}{\pgname_2}}
  }
\and
\inferrule*[Left=List-rec]
  {
    (\pcname_1 \land \prename) \implies \pcname_2
    \and
    \pcname_2[a_0 \mapsto \mcl{Cons(}h\mcl{, }t\mcl{)}] \implies \pcname_2[a_0 \mapsto t]
    \\
    \br{\seqa}{\pcname_2}{\phi[a_0 \mapsto \mcl{Nil}]}{\seqx} \vdash \pg{\btrue}{\pgname_1}
    \\
    \br{\seqr \splus h \splus t \splus \seqa}
       {\pcname_2[a_0 \mapsto \mcl{Cons(}h\mcl{, }t\mcl{)}] \land \phi[a_0 \mapsto t,\seqx \mapsto \seqr]}
       {\phi[a_0 \mapsto \mcl{Cons(}h\mcl{, }t\mcl{)}]}
       {\seqx}
       \vdash \pg{\btrue}{\pgname_2}
  }
  {
    \br{a_0 \splus \seqa}{\pcname_1}{\phi}{\seqx} \vdash \pg{\prename}{\mcl{rec(}a_0\mcl{, }\seqa\mcl{)}}
  }
\end{mathpar}
\caption{Selected synthesis inference rules.}
\label{fig:syninfrules}
\end{figure*}

Building on our correctness criteria for synthesis solutions, we now describe
\emph{inference rules} for synthesis. Such rules describe relations between
synthesis problems, capturing how some problems can be solved by reduction to
others. We have shown in previous work how to design a set of rules to ensure
\emph{completeness} of synthesis for a well-specified class of formulas, e.g.
integer linear arithmetic relations
\cite{KuncakETAL10CompleteFunctionalSynthesis} or simple term algebras
\cite{JacobsKuncakSuter13ReductionsSynthesisProcedures}. In the interest of
remaining self-contained, we shortly describe some generic rules. We then
proceed to presenting inference rules which allowed us to derive synthesis
solutions to problem that go beyond such decidable domains.

The validity of each rule can be established independently from its
instantiations, or from the context in which they it is used. This in turn
guarantees that the programs obtained by successive applications of validated
rules are correct by construction.

\paragraph{Generic reductions.}
As a first example, consider the rule \textsc{One-point} in
Figure~\ref{fig:syninfrules}. It intuitively reads as follows; ``if the
predicate of a synthesis problem contains a top-level atom of the form $x_0 =
t$, where $x_0$ is an output variable not appearing in the term $t$, then we
can solve a simpler problem where $t$ is substituted for $x_0$, obtain a
solution $\pg{\prename}{\pgname}$ and reconstruct a solution for the original
one by first computing the value for $t$ and then assigning as the result for
$x_0$''.

Another, perhaps simpler, example is given by \textsc{Ground} in
Figure~\ref{fig:syninfrules}. This rule simply states that if a synthesis
problem does not refer to any input variable, then it can be treated as a
satisfiability problem: any model for the predicate $\phi$ can then be used as
a ground solution term for $\seqx$.

\paragraph{Conditionals.}
The rules we have seen so far generate straight-line, unconditional
expressions. In order to synthesize programs that include conditional
expressions, we need rules such as \textsc{Case-split} in
Figure~\ref{fig:syninfrules}. The intuition behind \textsc{Case-split} is that
a disjunction in the synthesis predicate can be handled by an if-then-else
expression in the synthesized code, and each subproblem (corresponding to
predicates $\phi_1$ and $\phi_2$ in the rule) can be treated separately.  As
one would expect, the precondition for the final program is obtained by taking
the disjunction of the preconditions for the subproblems. This matches the
intuition that the disjunctive predicate should be realizable if and only if
one of its disjunct is. Note as well that even though the disjunction is
symmetrical, in the final program we necessarily privilege one branch over the
other one. This has the interesting side-effect that we can, as shown in the
rule, add the negation of the precondition $\prename_1$ to the path condition
of the second problem. This has the potential of triggering simplifications in
the solution of $\phi_2$. An extreme case being when the first precondition is
{\btrue} and the ``else'' branch becomes unreachable.

The \textsc{Case-split} rule as we presented it applies to disjunctions in
synthesis predicates. We should note that it is sometimes desirable to
explicitly introduce such disjunctions. For instance, our system includes rules
to introduce branching on the equality of two variables, to perform case
analysis on the types of variables (pattern-matching), etc. These rules can be
thought of as introducing first a disjunct, e.g. $a = b \lor a \neq b$, then
applying \textsc{Case-split}.

\paragraph{Recursion Schemas.}
We now show an example of an inference rule that produces a recursive function.
A common paradigm in functional programming is to perform a computation by
recursively traversing a structure. The rule \textsc{List-rec} captures one
particular form of such a traversal for the \cl{List} recursive type used in
the examples of Section~\ref{sec:examples}. The goal of the rule is to derive a
solution consists of a single invocation to a recursive function \cl{rec}. The
recursive function has the following form:
\begin{lstlisting}
def rec($a_0$, $\seqa$) = {
  require($\pcname_2$)
  $a_0$ match {
    case Nil $\RA$ $\pgname_1$
    case Cons($h$, $t$) $\RA$
      val $\seqr$ = rec($t$, $\seqa$)
      $\pgname_2$      
  }
} ensuring($\seqr$ $\RA$ $\phi[\seqx\mapsto\seqr]$)
\end{lstlisting}
where $a_0$ is of type \cl{List}. The function iterates over the list $a_0$
while preserving the rest of the input variables (the environment) $\seqa$.
Observe that its postcondition corresponds exactly to the synthesis predicate
of the original problem. We now go over the premises of the rule in detail:
\begin{itemize}
  \item The condition $(\pcname_1 \land P) \implies \pcname_2$ is necessary to
ensure that the initial call to \cl{rec} in the final program will
satisfy its precondition.
  \item The condition $\pcname_2[a_0 \mapsto \mcl{Cons(}h\mcl{, }t\mcl{)}] \implies \pcname_2[a_0 \mapsto t]$
states that the precondition of \cl{rec} should be inductive, i.e. whenever it
holds for a list, it should also hold for its tail. This is necessary to ensure
that the recursive call will satisfy the precondition.
  \item The subproblem $\br{\seqa}{\pcname_2}{\phi[a_0 \mapsto \mcl{Nil}]}{\seqx}$
corresponds to the base case (\cl{Nil}), and thus does not
contain the input variable $a_0$.
  \item The final subproblem is the most interesting, and corresponds to the
case where $a_0$ is a \cl{Cons}, represented by the fresh input variables $h$
and $t$. Because the recursive structure is fixed, we can readily represent the
result of the invocation \cl{rec($t$, $\seqa$)} by another fresh variable $r$.
We can assume that the postcondition of \cl{rec} holds for that particular
call, which we represent in the path condition as $\phi[a_0 \mapsto t,\seqx
\mapsto \seqr]$. The rest of the problem is obtained by substituting $a_0$ for
$\mcl{Cons(}h\mcl{, }t\mcl{)}$ in the path condition and in the synthesis
predicate.
\end{itemize}

\section{Exploring the Space of Subproblems}

In the previous section, we described a general formal framework in which we
can describe what constitutes a synthesis problem and a solution. In
particular, we have shown how synthesis rules decompose synthesis problems into
subproblems. In this section, we describe how we automatically search across
rule instantiations to derive a complete solution to a problem.

Inference rules are non-deterministic by nature. They justify the correctness
of a solution, but do not by themselves describe how one finds that solution.
Our search for a solution alternates between considering 1) which rules apply
to given problems, and 2) which subproblems are generated by rule
instantiations.

The task of finding rules that apply to a problem intuitively correspond to
finding an inference rule whose conclusion matches the structure of a problem.
For instance, to apply \textsc{Ground}, the problem needs to mention only
output variables. Similarly, to apply \textsc{List-rec} to a problem, it needs
to contain at least one input variable of type \cl{List}.

Computing the subproblems resulting from the application of a rule is in
general straightforward, as they correspond to problems appearing in its
premise. The \textsc{Ground} rule, for instance, generates no subproblem, while
\textsc{List-rec} generates two.

\paragraph{\andor search.}
To solve one problem, it suffices to find a complete derivation from \emph{one}
rule application to that problem. However, to fully apply a rule, we need to
solve \emph{all} generated subproblems. This corresponds to searching for a
closed branch in an \andor tree \cite{MartelliMontanari73AdditiveAndorGraphs}.

\begin{figure}
\centerline{\includegraphics[scale=0.75]{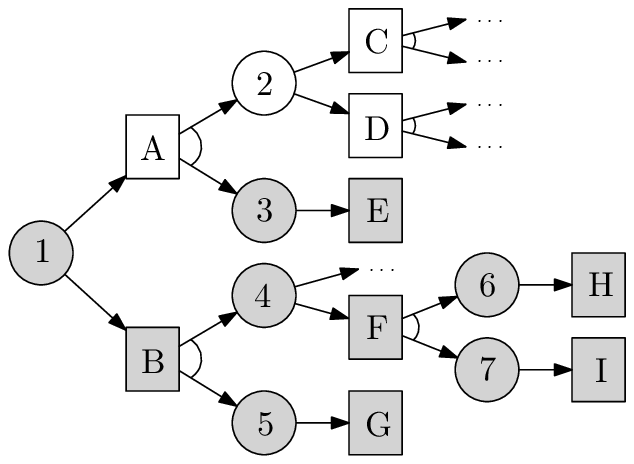}}
\vspace*{2ex}
\caption{An \andor search tree used to illustrate our search mechanism. 
Circles are \textsc{or} nodes and represent
problems, while boxes are \text{and} nodes and represent our rule applications.
Nodes in grey are closed (solved).}
\label{fig:andortree}
\end{figure}
We now describe the expansion of such a tree using an example.  Consider the
problem of removing a given element \cl{e} from a list \cl{a}.  In our logical
notation --using $\alpha$ as an abbreviation for \mcl{content}-- the problem
is:
\[
  \br{a \splus e}{\btrue}{\alpha(x) = \alpha(a) \setminus \{ e \}}{x}
\]
We denote this problem by $1$ in the tree of Figure~\ref{fig:andortree}. While
we haven't given an exhaustive list of all rules used in our system, it is fair
to assume that more than one can apply to this problem. For instance, we could
case-split on the type of $a$, or apply \textsc{List-rec} to $a$. We represent
these two options by A and B respectively in the tree.

Following the option B and applying \textsc{List-rec} with the path condition
$\pcname_2 \equiv \btrue$ trivially satisfies the first two premises of the
rules, and generates two new problems ($5$ and $6$). Problem $5$ is:
\[
  \br{e}{\btrue}{\alpha(x) = \alpha(\mcl{Nil}) \setminus \{ e \}}{x}
\]
where the predicate simplifies to $\alpha(x) = \emptyset$. This makes it
possible to apply the \textsc{Ground} rule (node G). This generates no
subproblem, and closes the subbranch with the solution solution
$\pg{\btrue}{\mcl{Nil}}$.  Problem $4$ has the form:
\begin{align*}
  \llbracket r \splus h \splus t \splus e \ \langle \alpha(r) = \alpha(t) \setminus \{ e \} \rhd \\
  \alpha(x) = \alpha(\mcl{Cons(}h\mcl{, }t\mcl{)}) \setminus \{ e \} \rangle \ x \rrbracket
\end{align*}
Among the many possible rule applications, we can choose to case-split on the
equality $h = e$ (node F). This generates two subproblems. Problem $6$
\begin{align*}
  \llbracket r \splus h \splus t \splus e \ \langle \alpha(r) = \alpha(t) \setminus \{ e \} \land e = h \rhd \\
  \alpha(x) = \alpha(\mcl{Cons(}h\mcl{, }t\mcl{)}) \setminus \{ e \} \rangle \ x \rrbracket
\end{align*}
and a similar problem $7$, where $e \neq h$ appears in the path condition
instead of $e = h$. Both subproblems can be solved by using a technique we will describe
in Section~\ref{sec:cegis} to derive a term satisfying the synthesis predicate,
effectively closing the complete branch from the root. The solutions for
problem $6$ and $7$ are $\pg{\btrue}{r}$ and
$\pg{\btrue}{\mcl{Cons(}h\mcl{,}r\mcl{)}}$ respectively. A complete reconstruction of the solution given by the branch in gray yields the program:
\begin{lstlisting}
def rec(a : List) : List = a match {
  case Nil $\RA$ Nil
  case Cons(h,t) $\RA$
    val r = rec(t)
    if(e $\EQ$ h)  r
    else Cons(h,r)
}
\end{lstlisting}

In the interest of space, we have only described the derivations that lead to
the search. In practice of course, not all correct steps are taken in the right
order. The interleaving of expansions of \textsc{and} and \textsc{or} nodes is
driven by the \emph{estimated cost} of problems and solutions.

\paragraph{Cost models.}
In order to drive the search, we assign to each problem and to each rule
application an estimated cost, which is supposed to under-approximate to actual
final cost of a closed branch. For \textsc{or} nodes (problems), the cost is
simply the minimum of all remaining viable children, while for \textsc{and}
nodes (rule applications) we take the sum of the cost of each children plus a
constant. That constant intuitively corresponds to the extra complexity
inherent to a particular rule.

A perfect measure for cost would be the running time of the corresponding
program. However, this is particularly hard to estimate, and valid
under-approximations would most likely be useless. We chose to measure program
size instead, as we expect it to be an reasonable proxy for complexity. We
measure the size of the program as the number of branches, weighted by their
proximity to the root. We found this to be have a positive influence on the
quality of solutions, as it discourages near-top-level branching.

Using this metric, the cost inherent to a rule application roughly corresponds
to the extra branches it introduces in the program. We use a standard algorithm
for searching for the best solution
\cite{MartelliMontanari73AdditiveAndorGraphs}, and the search thus always
focuses on the current most promising solution. In our example in
Figure~\ref{fig:andortree}, we could imagine that after the case split at F,
the B branch temporarily became less attractive. The search then focuses for a
while on the A branch, until expansion on that side (for instance by
case-splitting on the type of the list) reached a point where the minimal
possible solution was worse than the B branch. We note that the complete search
takes about two seconds.

\paragraph{Anytime synthesis.}
Because we maintain the search tree and know the current minimal solution at
all times, we can stop the synthesis at any time and obtain a partial program
that is likely to be good. This option is available in our implementation, both
from the console mode and the web interface. In such cases, \leon will return a
program containing new invocations of \cl{choose} corresponding to the open
subproblems.

%
%

\section{Symbolic Term Exploration}
\label{sec:cegis}

In previous sections, we have introduced the notion of synthesis inference
rules, and described how to search over rule applications that generate
subproblems. In this section, we describe one of our most important rules,
which is responsible for closing most of the branches in search trees. We call
it Symbolic Term Exploration (\ourcegis).

The core idea behind \ourcegis is to symbolically represent many possible terms
(programs), and to iteratively prune them out using counterexamples and test
case generation until either 1) a valid term is proved to solve the synthesis
problem or 2) all programs in the search spaces have been shown to be
inadequate. Since we already have rules that take care of introducing
branching constructs or recursive functions, we focus \ourcegis on the search
for terms consisting only of constructors and calls to existing functions.

\paragraph{Recursive generators.}
We start from a universal non-deterministic
program that captures all the (deterministic) programs which we wish to
consider as potential solutions. We then try to resolve the non-deterministic
choices in such a way that the program realizes the desired property. Resolving
the choices consists in fixing some values in the program, which we achieve by
running a counterexample driven search.

We describe our non-deterministic programs as a set of
recursive non-deterministic \emph{generators}. Intuitively, a generator for a
given type is a program that produces arbitrary values of that type. For
instance, a generator for positive integers could be given by:
\begin{lstlisting}
def genInt() : Int = if($\star$) 0 else (1 + genInt())
\end{lstlisting}
where $\star$ represents a non-deterministic boolean value. Similarly
a non-deterministic generator for the \cl{List} type could take the form:
\begin{lstlisting}
def genList() : List = if($\star$) Nil else Cons(genInt(), genList())
\end{lstlisting}

It is not required that generators can produce \emph{every} value for a given
type; we could hypothesize for instance that our synthesis solutions will only
need some very specific constants, such as $0$, $1$ or $-1$.  What is more
likely is that our synthesis solutions will need to use input variables and
existing functions. Our generators therefore typically include variables of the
proper type that are accessible in the synthesis environment. Taking these
remarks into account, if \cl{a} and \cl{b} are integer variables in the scope,
and \cl{f} is a function from \cl{Int} to \cl{Int}, a typical generator for
integers would be:
\begin{lstlisting}
def genInt() : Int = if($\star$) 0 else if($\star$) 1 else if($\star$) -1
        else if($\star$) a else if($\star$) b else f(genInt())
\end{lstlisting}

\paragraph{From generators to formulas.} Generators can in principle be any
function with unresolved non-deterministic choices. For the sake of the
presentation, we assume that they are ``flat'', that is, they consist of a
top-level non-deterministic choice between $n$ alternatives. (Note that the
examples given above all have this form.)

Encoding a generator into an SMT term is straightforward: introduce for each
invocation of a generator an uninterpreted constant $c$ of the proper type, and
for each non-deterministic choice as many boolean variables $\seq{b}$ as there are
alternatives. Encode that exactly one of the $\seq{b}$ variables must be true, and
constrain the value of $c$ using the $\seq{b}$ variables.


Recursive invocations of generators can be handled similarly, by inserting
another $c$ variable to represent their value and constraining it
appropriately. Naturally, these recursive instantiations must stop at some
point: we then speak of an \emph{instantiation depth}. As an example, the
encoding of the \cl{genList} generator above with an instantiation depth of $1$
and assuming that \cl{genInt} generates $0$ or $a$ is:
\begin{align*}
          & (b_1 \lor b_2) \land (\neg b_1 \lor \neg b_2) \\
  \land~~ & b_1 \RA c_1 = \mcl{Nil} 
  \land     b_2 \RA c_1 = \mcl{Cons(}c_2\mcl{, }c_3\mcl{)} \\
  \land~~ & (b_3 \lor b_4) \land (\neg b_3 \lor \neg b_4) \\
  \land~~ & b_3 \RA c_2 = 0
  \land     b_4 \RA c_2 = a \\
  \land~~ & (b_5 \lor b_6) \land (\neg b_5 \lor \neg b_6) \\
  \land~~ & b_5 \RA c_3 = \mcl{Nil}
  \land     b_6 \RA c_3 = \mcl{Cons(}c_4\mcl{, }c_5\mcl{)} \\
  \land~~ & \neg b_6
\end{align*}
The clauses encode the following possible values for $c_1$: \cl{Nil},
\cl{Cons(0, Nil)} and \cl{Cons(a, Nil)}. Note the constraint $\neg b_6$ which
encodes the instantiation depth of $1$, by preventing the values beyond that
depth (namely $c_4$ and $c_5$) to participate in the expression.

For a given instantiation depth, a valuation for the $\seq{b}$ variables encodes a
determinization of the generators, and as a consequence a program. We solve for
such a program by running a refinement loop.

\paragraph{Refinement loop: discovering programs.} Consider a synthesis problem
$\br{\seqa}{\pcname}{\phi}{\seqx}$, where we speculate that a generator for the
types of $\seqx$ can produce a program that realizes $\phi$. We start by
encoding the non-deterministic execution of the generator for a fixed
instantiation depth (typically, we start with $0$). Using this encoding, the
problem has the form:
\begin{equation}
\label{eqn:ourcegisbase}
  \phi \land B(\seqa, \seq{b}, \seq{c}) \land C(\seq{c}, \seqx)
\end{equation}
where $\phi$ is the synthesis problem, $B$ is the set of clauses obtained by
encoding the execution of the generator and $C$ is a set of equalities tying
$\seqx$ to a subset of the $\seq{c}$ variables. Note that by construction,
the values for $\seq{c}$ (and therefore for $\seqx$) are uniquely determined
when $\seqa$ and $\seq{b}$ are fixed.

We start by finding values for $\seqa$ and $\seq{b}$ such that
\eqref{eqn:ourcegisbase} holds. If no such values exist, then our generators at
the given instantiation depth are not expressive enough to encode a solution to
the problem. Otherwise, we extract for the model the values $\seq{b}_0$.
They describe a candidate program, which we put to the test.

\paragraph{Refinement loop: falsifying programs.}
We search for a solution to the problem:
\begin{equation}
\label{eqn:ourcegisrefute}
  \neg \phi \land B(\seqa, \seq{b}_0, \seq{c}) \land C(\seq{c}, \seqx)
\end{equation}
Note that $\seq{b}_0$ are constants, and that $\seq{c}$ and $\seqx$ are
therefore uniquely determined by $\seqa$ this intuitively comes from the fact
that $\seq{b}_0$ encodes a deterministic program, that $\seq{c}$ encodes
intermediate values in the execution of that program, and that $\seqx$
encodes the result. With this in mind, it becomes clear that we are really solving
for $\seqa$.

If no such $\seqa$ exist, then we have found a program that realizes $\phi$
and we are done. If on the other hand we can find $\seq{a}_0$, then this
constitutes an input that witnesses that our program does not meet the
specification. In this case, we can discard the program by asserting $\neg
\bigwedge \seq{b}$, and going back to \eqref{eqn:ourcegisbase}.

Eventually, because the set of possible assignments to $\seq{b}$ is finite (for
a given instantiation depth) this terminates. If we have not found a program,
we can increase the instantiation depth and try again. When the maximal depth
is reached, we give up.

\paragraph{Filtering with concrete execution.}
While termination is in principle guaranteed just by successive elimination of
programs in the refinement loop, the formula encoding the non-deterministic
program typically grows exponentially as instantiation depth increases.
As the number of programs grows, the difficulty for the solver to satisfy
\eqref{eqn:ourcegisbase} or \eqref{eqn:ourcegisrefute} also increases.  As an
alternative to symbolic elimination, we can often use concrete execution on a
set of inputs to rule out many programs. We rely on \leon's capability for
small model finding (see Section~\ref{sec:leonverifier}) to generate inputs
that satisfy the path condition. We then use on-the-fly code generation to
compile the symbolic program into a function that takes as arguments the input
variables as well as a boolean array encoding the non-deterministic choices.
This allows us to rapidly discard hundreds or even thousand of programs.
Whenever the change is substantial, we regenerate a new formula for
\eqref{eqn:ourcegisbase} with much fewer boolean variables and continue from
there. Note that very often, when \ourcegis is applied to a problem it cannot
solve, concrete execution rules out all programs in a fraction of a second and
symbolic reasoning is never applied.

\section{Type-Driven Counterexample-Guided Synthesis with Condition Abduction}
\label{sec:insynth}

Our second larger rule focuses on synthesizing recursive functions that satisfy
a given specification.
We assume that we are given a function header and a postcondition, and that
we aim to synthesize a recursive function body.
Note that the expression must be 1) a
well-typed term with respect to the context of the program and 2) 
valid according to the imposed formal specification.
Therefore, an approach to solve this kind of synthesis problems could be based
on searching the space of all expressions that can be built from all
declarations visible at the corresponding place in the program, i.e. in the
scope of \cl{choose}, while limiting attention to those that type-check,
have the desired type, and satisfy the given formal specification.

An obvious drawback of such approach is 
that, unless the process 
is carefully guided, the search becomes unfeasible
due to search space explosion.
In practice we indeed found that trivial generate-and-test strategies scale
poorly with the number of visible declarations and the search becomes
practically unfeasible even for small programs.




\subsection{Condition Abduction}


Our idea for guiding the search and incremental construction of
correct expressions comes from the area of abductive reasoning
\cite{Josephson94AbductiveInferenceComputationPhilosophyTechnology,
KakasKowalskiToni92AbductiveLogicProgramming}.
Abductive reasoning, sometimes also called ``inference to the best
explanation'', is a method of reasoning in which one chooses a hypothesis that would
explain the observed evidence in the best way.

The motivation behind the approach to applying abductive reasoning to program
synthesis
comes from examining implementations of
practical purely functional, recursive algorithms.
The key observation is that recursive functional algorithms share a similar
pattern.
They implement behaviour through a combination of case analysis with
control flow expressions (e.g. if-then-else) and recursive calls.
This pattern is encoded with
a branching control flow expression
that partitions the space of input values such that each branch represents a
correct implementation for a certain partition.
Such partitions are defined by conditions that guard branches in the control
flow.

This allows synthesizing branches separately by searching for implementations
that evaluate correctly only for certain inputs while restricting the search space.
Rather than speculatively applying \textsc{Case-split} rule to obtain
subproblems and finding solutions for each branch by case analysis (as
described in Section \ref{sec:framework}), this idea applies a similar strategy
in the reverse order -- getting a candidate program and searching for a condition that
would make it correct.
Thus, the idea of abductive reasoning can be applied to guess the
condition that defines a valid partition,
i.e. ``abduce'' the explanation for a partial implementation, with
respect to a given candidate program.
The rule progressively applies this technique 
and enables effective search and construction of a control
flow expression that represents a correct implementation for more and more input
cases, eventually constructing an expression that is a solution to the synthesis problem.

\subsection{The Algorithm Used in the Rule}
\label{subsubsec:the_synthesis_algorithm}

Based on these observations, we present
our rule that employs a new technique for guiding the search
with ranking and filtering based on counterexamples, as well as
 constructing expressions
from partially correct implementations.
It is presented in Algorithm \ref{alg:synthesize_correct_program}.

\setlength{\textfloatsep}{1,38em}
\begin{algorithm}[ht]
\caption{Synthesis with condition abduction}
\label{alg:synthesize_correct_program}
\begin{algorithmic}[1]

\Require{path condition $\pcname$, predicate $\phi$, a collection of\hspace{1em}
expressions $s$} {\hfill\parbox[t]{.7\linewidth}{\Comment{synthesis problem $\br{\seqa}{\pcname}{\phi}{\seqx}$}}}

\State $p' = true$ \Comment{maintain the current partition}
\State $sol = (\lambda x. x)$ \Comment{maintain a partial solution}
\State $\mathcal{M} =$ \textsc{SampleModels($\seq a$)} \Comment{set of example
models}
\Repeat
\State get a set of expressions $E$ from $s$ \Comment{candidates}
\ForAll{$e$ in $E$} \Comment{count passed examples $p_e$ for $e$}
\State $p_e = | \{ m \in \mathcal{M} \mid e(m) \mbox{ is correct}\} |$
\Comment{evaluate}
\EndFor 
\State $\seq{r} = \arg\max_{\;e \in E\;} p_e$ \Comment{the highest ranked
expression}
\If{solution $\pg{\pcname \wedge p'}{\seq{r}}$ is valid}
\State \Return $\pg{\pcname}{(sol\;\seq{r})}$ \Comment{a solution is found}
\Else 
\State{extract new counterexample model $m$}
\State{$\mathcal{M} = \mathcal{M}
\cup m$} \Comment{accumulate examples}
\State{$c =$ \Call{BranchSyn}{$\seq{r}, p, q, s$}}
 \Comment{call Algorithm \ref{alg:synthesize_correct_expressions}}
\If{$c \neq$ \textsc{false}} \Comment{a branch is synthesized}
\State $sol = (\lambda x.$ $(sol\;($if $c$ then $\seq{r}$ else $x)))$ 
\State{$p' = p' \wedge \neg c$} \Comment{update current partition}
\EndIf
\EndIf
\Until{$s$ is not empty}
\end{algorithmic}
\end{algorithm}

The algorithm applies the idea of abducing conditions to progressively
synthesize and verify branches of a correct implementation for an expanding
partition of inputs.
The input to the algorithm is a path condition $\pcname$, a
predicate $\phi$ (defined by synthesis problem
$\br{\seqa}{\pcname}{\phi}{\seqx}$), and a collection of expressions $s$.

Condition $p'$ defines which inputs are left to consider at any given point in
the algorithm; these are the inputs that belong to the current partition.
The initial value of $p'$ is $true$, so the algorithm starts with a
partition that covers the whole initial input space constrained only by the
path condition $\pcname$.
Let $p_1, \ldots, p_k$, where $k > 0$, be conditions abduced up to a
certain point in the algorithm.
Then $p'$ represents the conjunction of negations of abduced conditions, i.e.
$p' = \neg p_1 \wedge \ldots \wedge \neg p_k$.
Together with the path condition, it defines the current partition which includes all
input values for which there is no condition abduced (nor correct implementation found).
Thus, the guard condition for the current partition is defined by $\pcname
\wedge p'$.
The algorithm maintains the partial solution  $sol$, encoded as a function.
$sol$ encodes an expression which is
correct for all input values that satisfy any of the abduced conditions and this
expression can be returned as a partial solution at any point.
Additionally, the algorithm accumulates example models in the set $\mathcal{M}$.
Ground term generator, described in Section \ref{sec:leonverifier},
is used to construct the initial set of models in $\mathcal{M}$.
To construct a model, for each variable in $\seq a$, the algorithm assigns
a value sampled from the ground term generator.
Note that more detailed discussion on how examples are used to
guide the search is deferred to Section \ref{subsec:organizzation_of_the_search}.

The algorithm repeats enumerating all possible expressions from the given
collection until it finds a solution.
In each iteration, a batch of expressions $E$ is enumerated and evaluated
on all models from $\mathcal{M}$.
The results of such evaluation are used to rank expressions from
$E$.
The algorithm considers the expression of the highest rank $\seq{r}$
as a candidate solution and checks it for validity.
If $\seq{r}$ represents a correct implementation for the current partition, i.e. if $\pg{\pcname \wedge
p'}{\seq{r}}$ is a valid solution,
then the expression needed to complete a valid control flow expression is found.
The algorithm returns it as solution for which $\br{\seqa}{\pcname}{\phi}{\seqx}
\vdash \pg{\pcname}{(sol\;\seq{r})}$ holds.
Otherwise, the algorithm extracts the counterexample model
$m$, adds it to the set $\mathcal{M}$, and continues by trying to synthesize a branch
with expression $\seq{r}$ (it does so by calling Algorithm
\ref{alg:synthesize_correct_expressions} which will be explained later).
If \textsc{BranchSyn} returns a valid branch condition, the algorithm updates the partial
solution to include the additional branch (thus extending extending the space of inputs
covered by the partial solution), and refines the current partition condition.
New partition condition reduces the synthesis to a subproblem, ensuring that the 
solution in the next iteration covers cases where $c$ does not hold.
The algorithm eventually, given the
appropriate terms from $s$, finds an expression that forms
a complete correct implementation for the synthesis problem.

\begin{algorithm}[ht]
\caption{Synthesize a branch}
\label{alg:synthesize_correct_expressions}
\begin{algorithmic}[1]

\Require{expression $\seq{r}$, condition $p'$, predicate $q$, and a\hspace{1em}
collection of expressions $s$} \Comment{passed from Algorithm \ref{alg:synthesize_correct_program}}
\Function{BranchSyn}{$\seq{r}, p', q, s$}

\State $\mathcal{M'} = \emptyset$ \Comment{set of accumulated counterexamples}

\State get a set of expressions $E'$ from $s$ \Comment{candidates}
\ForAll{$c$ in $E'$}
\If{\textbf{for each} model $m$ in $\mathcal{M'}$, $c(m) = false$}
\If{ solution $\pg{\pcname \wedge c}{\seq{r}}$ is valid }
\State \Return $c$ \Comment{a condition is abduced}
\Else 
\State{extract the new counterexample model $m$}
\State{$\mathcal{M'} = \mathcal{M'} \cup m$} \Comment{accumulate
counterexamples}
\EndIf
\EndIf
\EndFor
\State{return \textsc{false}} \Comment{no condition is found}

\EndFunction
\end{algorithmic}
\end{algorithm}


Algorithm \ref{alg:synthesize_correct_expressions} tries to
synthesize a new branch by abducing a valid branch condition $c$.
It does this by enumerating a set of expressions $E'$ from $s$ and checking
whether it can find a valid condition expression, that would guard a partition for which
the candidate expression $\seq{r}$ is correct.
The algorithm accumulates counterexamples models in $\mathcal{M'}$ and
considers a candidate expression $c$ only if it prevents all accumulated
counterexamples.
The algorithm checks this by evaluating $c$ on $m$, i.e. $c(m)$, for each accumulated
counterexample $m$.
If a candidate expression $c$ is not filtered out, the algorithm checks if $c$
represents a valid branch condition, i.e. whether $\pg{\pcname \wedge
c}{\seq{r}}$ is a valid solution.
If yes, the algorithm returns $c$ which, together with $\seq{r}$, comprises a valid branch in
the solution to $\br{\seqa}{\pcname}{\phi}{\seqx}$.
Otherwise, it adds a new counterexample model to $\mathcal{M'}$ and continues with the
search.
If no valid condition is in $E'$, the algorithm returns
\textsc{false}.

\subsection{Organization of the Search}
\label{subsec:organizzation_of_the_search}

For getting the collection of expressions $s$,
the rule uses term generators that generate
all well typed terms according to type constraints derived from the
context of a program \cite{Gvero13CompleteCompletionUsingTypesAndWeights,
Kuraj13InteractiveCodeGeneration}.
This has the advantage of initial search space restriction inherent to the
generator that limits enumerated expressions only to those that are well typed.
The completeness property of such generators ensures systematic enumeration of
all candidate solutions that are defined by the set of given type constraints.
For verification, the rule uses 
the {\leon} verifier,
that allows checking validity of expressions that are supported by
the underlying theories and obtaining counterexample models.

The context of the algorithm as a rule in the \leon synthesis framework imposes
limits on the portion of search space explored by each rule instantiation.
This allows incremental and systematic progress in search space exploration and,
due to the mixture with other synthesis rules, 
offers benefits in both expressiveness and performance of synthesis.
The rule offers flexibility in adjusting necessary parameters and thus a
fine-grain control over the search - for our experiments, the size of candidate
sets of expressions enumerated in each iteration $n$ is 50 (and is doubled in
each iteration) and 20, in the case of Algorithm \ref{alg:synthesize_correct_program} and
\ref{alg:synthesize_correct_expressions}, respectively.

\paragraph{Using (counter-)examples.}


%

A technique that brings significant performance improvements when
dealing with large search spaces is guiding the search and even avoid considering
candidate expressions according to the information from examples generated during synthesis.
As described earlier, after checking an unsatisfiable formula, the rule queries
\leon for the witness model and accumulates examples that are used to narrow
down the search space.

Algorithm \ref{alg:synthesize_correct_expressions} uses accumulated
counterexamples to filter out unnecessary candidate expressions when synthesizing a branch.
It makes sense to consider a candidate expression for a branch condition, $c$,
for a check whether $c$ makes $\seq{r}$ a correct implementation,
only if $c$
prevents all accumulated counterexamples that already witnessed unsatisfiability
of the correctness formula for $\seq{r}$, i.e. if
$\forall m \in \mathcal{M'}. \;c \rightarrow \neg m$.
Otherwise, if $\exists m \in \mathcal{M'}. \neg(c \rightarrow \neg m)$, then $m$
is a valid counterexample to the verification of $\pg{\pcname \wedge
c}{\seq{r}}$.
This effectively guides the search by the results of previous
verification failures while filtering out candidates before more expensive verification check are made.

Algorithm \ref{alg:synthesize_correct_program} uses accumulated models to
quickly test and rank expressions 
by evaluating models according to the specification.
The current set of candidate expressions $E$ is evaluated on the set
of accumulated examples $\mathcal{M}$ and results of such evaluation are used to
rank the candidates.
We call an evaluation of a candidate $e$ on a model $m$ correct, if $m$
satisfies path condition $\pcname$ and the result of the evaluation
satisfies given predicate $q$.
The algorithm counts the number of correct evaluations, ranks
the candidates accordingly and considers only the candidate of the highest rank.
The rationale is that the more correct evaluations, the more likely the
candidate represents a correct implementation for some partition of inputs.
Note that evaluation results may be used only for ranking but not for filtering,
because each candidate may
represent a correct implementation for a certain partition of inputs, thus
incorrect evaluations are expected even for valid candidates.
Since the evaluation amounts to executing the specification
this technique is efficient in guiding the search toward correct correct
implementations while avoiding unnecessary verification checks. 

  \newcommand{\bname}[1]{{\small \texttt{#1}}}
\begin{figure}
    \begin{center}
    \begin{tabular}{@{}l@{\,}r@{\,}c@{\,}c@{\ }r@{\,}r}
        Operation                           & Syn   & Size  & Calls  & sec.  & Proved      \\
        \hline
        \bname{List.Insert}                &    $\surd$    & 3     & 0 &     0.3   &  $\surd$    \\
        \bname{List.Delete}                &    $\surd$    & 19    & 1 &     2.0   &  $\surd$   \\
        \bname{List.Union}                 &    $\surd$    & 12    & 1 &     2.0   &  $\surd$   \\
        \bname{List.Diff}                  &    $\surd$    & 12    & 2 &     7.0   &  $\surd$   \\
        \bname{List.Split}                 &    $\surd$    & 27    & 1 &     2.0   &  $\surd$   \\
        \hline
        \bname{SortedList.Insert}          &    $\surd$    & 34    & 1 &     8.9   &  $\surd$   \\
        \bname{SortedList.InsertAlways}    &    $\surd$    & 36    & 1 &     12.5  &  $\surd$   \\
        \bname{SortedList.Delete}          &    $\surd$    & 23    & 1 &     8.7   &            \\
        \bname{SortedList.Union}           &    $\surd$    & 19    & 2 &     5.0   &  $\surd$   \\
        \bname{SortedList.Diff}            &    $\surd$    & 13    & 2 &     6.8   &            \\
        \bname{SortedList.InsertionSort}   &    $\surd$    & 10    & 2 &     5.1   &  $\surd$   \\
        \bname{SortedList.MergeSort}       &    $\surd$    & 11    & 4 &    87.7   &  $\surd$   \\
        \hline
        \bname{StrictSortedList.Insert}    &    $\surd$    & 34    & 1 &     9.9   &  $\surd$   \\
        \bname{StrictSortedList.Delete}    &    $\surd$    & 21    & 1 &    16.1   &            \\
        \bname{StrictSortedList.Union}     &    $\surd$    & 19    & 2 &     4.1   &  $\surd$   \\
        \hline
        \bname{UnaryNumerals.Add}         &    $\surd$    & 11    & 1 &     1.6   &  $\surd$   \\
        \bname{UnaryNumerals.Distinct}    &    $\surd$    & 12    & 0 &     1.9   &  $\surd$   \\
        \bname{UnaryNumerals.Mult}        &    $\surd$    & 12    & 1 &     2.5   &  $\surd$   \\
    \end{tabular}
    \end{center}
\caption{We consider a problem as synthesized if the solution generated is
correct after manual inspection. For each generated program, we provide the
size of its syntax tree and the number of function calls it contains. Proved
problems are those for which the synthesized program can be automatically
proven to match its specification.}
\label{fig:evaluation}
\end{figure}

\section{Implementation and Results}

We have implemented these techniques in \leon, a system for verification and synthesis of
functional program, thus extending it from the state described in Section~\ref{sec:leonverifier}.
Our implementation and the online interface are available from
\url{http://lara.epfl.ch/leon/}.

The front end to \leon is the standard Scala compiler (for Scala 2.9). Scala
compiler performs type checking and tasks such as the expansion of implicit
conversions, from which \leon directly benefits. \leon programs also execute as
valid Scala programs. \leon checks that the syntax trees produced conform to
the subset that it expects and then performs verification and synthesis.

We have developed several interfaces for \leon. Leon can be invoked as a batch command-line
tool that accepts verification and synthesis tasks and outputs the results of the requested tasks.
If desired, there is also a console mode that allows applying synthesis rules in a step-by-step
fashion and is useful for debugging purposes.

To facilitate interactive experiments and the use of the system in teaching, we have also developed
an interface that executes in the web browser, using the Play framework of Scala as well as 
JavaScript editors. Our browser-based interface supports continuous compilation of Scala code,
allows verifying individual functions with a single keystroke or click, as well as synthesizing
any given \cl{choose} expression. In cases when the synthesis process is interrupted, the synthesizer
can generate a partial solution that contains a program with further occurrences of the \cl{choose}
statement.

\subsection{Results}

In order to evaluate our system, we developed benchmarks with reusable
abstraction functions. These abstraction functions allow for a concise
specification of each operation without requiring any insight on its resulting
implementation. It is interesting to notice that these functions generally
abstract any structural invariant inherent to the underlying data-structure.
For instance, the synthesis of
\begin{lstlisting}
def add(a: Num, b: Num) = choose {
    (res: Num) $\RA$ value(r) $\EQ$ value(a) + value(b)
}
\end{lstlisting}
would result in vastly different programs depending on the implementation of
\lstinline/Number/.

Our set of benchmarks displayed in Figure~\ref{fig:evaluation} covers the
synthesis of various operations over custom data-structures with invariants,
specified through the lens of abstraction functions. These benchmarks use
specifications with are both easy to understand and much shorter than resulting
programs (except in trivial cases). We believe these are key factors in the
evaluation of any synthesis procedure. 
The definitions and specifications of
all the benchmarks can be found in appendix.

Synthesis is performed in order, meaning that an operation will be able to
reuse all previously synthesized ones, thus mimicking the usual development
process.

We can see in Figure~\ref{fig:evaluation} the list of programs we successfully
synthesized. Each synthesized program has been manually validated to be a solution
that a programmer might expect. Our
system typically also proves automatically that the resulting program
matches the specification for all inputs. In certain cases, the lack of inductive
invariants prevents such fully-automated proof, which is a limitation of our verifier. Note
that we stop verification after a timeout of 3 seconds.

In almost all cases, the synthesis succeeds sufficiently fast for a reasonable
interactive experience.

\section{Related Work}


Our approach blends deductive
synthesis \cite{MannaWaldinger71TowardAutomaticProgramSynthesis,
  MannaWaldinger80DeductiveApproachtoProgramSynthesis,DBLP:conf/vstte/Smith05}, 
which incorporates transformation of
specifications, inductive reasoning, recursion schemes and termination checking, with
modern SMT techniques and constraint
solving for executable constraints. As one of our subroutines we include complete
functional synthesis for integer linear arithmetic \cite{KuncakETAL12SoftwareSynthesisProcedures} 
and extend it with complete functional synthesis for algebraic data types 
\cite{Suter12ProgrammingSpecifications,JacobsKuncakSuter13ReductionsSynthesisProcedures}.
This gives us building blocks for synthesis of recursion-free code. To synthesize
recursive code we build on and further advance the counterexample-guided approach to synthesis
\cite{SolarLezamaETAL06Combinatorialsketchingforfiniteprograms}.

\paragraph{Deductive synthesis frameworks.}
Early work on synthesis
\cite{MannaWaldinger71TowardAutomaticProgramSynthesis,
  MannaWaldinger80DeductiveApproachtoProgramSynthesis}
focused on synthesis using expressive and undecidable
logics, such as first-order logic and logic containing
the induction principle.

Programming by refinement has been popularized as a manual
activity \cite{DBLP:journals/cacm/Wirth83,BackWright98RefinementCalculus}.
Interactive tools have been developed to support such techniques in HOL
\cite{ButlerETAL97RefinementCalculator}.
A recent example of deductive synthesis and refinement is
the Specware
system from Kesterel \cite{DBLP:conf/vstte/Smith05}.
We were not able to use the system first-hand due to its availability policy,
but it appears to favor expressive power and control, whereas
we favor automation.

A combination of automated and interactive development is
analogous to the use of automation in interactive theorem
provers, such as Isabelle
\cite{NipkowETAL02IsabelleHOL}. However, whereas in
verification it is typically the case that the program is
available, the emphasis here is on constructing the program
itself, starting from specifications.

Work on synthesis from specifications
\cite{SrivastavaETAL10FromProgramVerificationtoProgramSynthesis}
resolves some of these difficulties by decoupling the
problem of inferring program control structure and the
problem of synthesizing the computation along the control
edges.  The work leverages verification
techniques that use both approximation and lattice theoretic
search along with decision procedures, but appears to require
more detailed information about the structure of the expected
solution than our approach.

\paragraph{Synthesis with input/output examples.}
One of the first works that addressed synthesis with examples and put inductive
synthesis on a firm theoretical foundation is
the one by Summers \cite{Summers77MethodologyLispProgramConstructionFromExamples}.
Subsequent work presents extensions of the classical approach to
induction of functional Lisp-programs \cite{KitzelmannSchmid06InductiveSynthesisFunctionalProgramsExplanationBased,Hofmann10IgorII}.
These extensions include synthesizing a set of equations (instead of just
one), multiple recursive calls and systematic introduction of parameters. Our current system lifts several restrictions of previous approaches by supproting
reasoning about arbitrary datatypes,
supporting multiple parameters in concrete and symbolic I/O examples, and
allowing nested recursive calls and user-defined
declarations. 

Inductive (logic) programming that explores automatic synthesis of (usually
recursive) programs from incomplete specifications, most often being
input/output examples
\cite{FlenerPartridge01InductiveProgramming,MuggletonRaedt94InductiveLogicProgrammingTheoryMethods},
influenced our work.
Recent work in the area of programming by demonstration has shown that synthesis
from examples can be effective in a variety of domains, such as spreadsheets 
\cite{SinghGulwani12SynthesizingNumberTransformationsInputOutputExamples}.
Advances in the field of SAT and SMT solvers inspired 
counter-example guided iterative synthesis \cite{SolarLezamaETAL06Combinatorialsketchingforfiniteprograms,GulwaniETAL11SynthesisLoopfreePrograms}, which can derive input and output examples from specifications.
Our tool uses and advances these techniques through two new
counterexample-guided synthesis approaches.

\paragraph{Synthesis based on finitization techniques.}
Program sketching has demonstrated the practicality of
program synthesis by focusing its use on particular domains
\cite{SolarLezamaETAL06Combinatorialsketchingforfiniteprograms,
  SolarLezamaETAL07Sketchingstencils,
  SolarLezamaETAL08Sketchingconcurrentdatastructures}.  The
algorithms employed in sketching are typically focused on
appropriately guided search over the syntax tree of the
synthesized program. The tool we presented shows one way
to move the ideas of sketching towards
infinite domains. In this generalization we leverage reasoning
about equations as much as SAT techniques. 

\paragraph{Reactive synthesis.}
Synthesis of reactive systems generates programs that run
forever and interact with the environment.  However, known
complete algorithms for reactive synthesis work with
finite-state systems
\cite{PnueliRosner89SynthesisofReactiveModule} or timed
systems
\cite{AsarinETAL95SymbolicControllerSynthesisforDiscrete}.
Such techniques have applications to control the behavior of
hardware and embedded systems or concurrent programs
\cite{VechevETAL09InferringSynchronizationunderLimitedObservability}.
These techniques usually take specifications in a fragment
of temporal logic \cite{PitermanETAL06SynthesisofReactiveDesigns} and have
resulted in tools that can synthesize useful hardware components
\cite{JobstmannETAL07Anzu,
JobstmannBloem06OptimizationsforLTLSynthesis}. 
Recently such synthesis techniques have been extended to repair that preserves
good behaviors \cite{EssenJobstmann13ProgramRepairRegret}, which is
related to our notion of partial programs that have remaining \cl{choose} statements.

\section{Conclusions and Analysis}

Software synthesis is a difficult problem but we believe it
can provide substantial help in software development. We
have presented a new framework for synthesis that combines
transformational and counterexample-guided approaches. Our
implemented system can synthesize and prove correct
functional programs that manipulate unbounded data
structures such as algebraic data types. We have used the
system to synthesized algorithms that manipulate list and
tree structures.  The algorithm can be combined with manual
transformations or run-time constraint solving to cover the
cases where static synthesis does not fully solve the
problem. Our current counterexample-guided synthesis steps are
domain-agnostic, while somewhat limits their scalability, so
we expect improved results using domain-specific generators,
such as the ones used in testing tools UDITA 
\cite{GligoricETAL10TestGenerationThroughProgramming} and 
Quickcheck \cite{ClaessenHughes00QuickCheck}.
Our framework leverages the state of the art SMT
solving technology and an effective mechanism for solving
certain classes of recursive functions. Thanks to this
technology, it was able to synthesize programs over unbounded
domains that are guaranteed to be correct for all inputs.

\acks
We thank Regis Blanc for his contribution to the Leon verification infrastructure.
We thank Tihomir Gvero and Ruzica Piskac for many discussions on synthesis.

\bibliographystyle{abbrvnat}
\bibliography{main}

\appendix
\newpage
\appendix
\section{Benchmarks Definitions}
\subsection{List}
\begin{lstlisting}
object ListBenchmark {
  sealed abstract class List
  case class Cons(head: Int, tail: List) extends List
  case object Nil extends List

  def size(l: List) : Int = (l match {
      case Nil $\RA$ 0
      case Cons(_, t) $\RA$ 1 + size(t)
  }) ensuring(res $\RA$ res $\SGE$ 0)

  def content(l: List): Set[Int] = l match {
    case Nil $\RA$ Set.empty[Int]
    case Cons(i, t) $\RA$ Set(i) $\PP$ content(t)
  }

  def abs(i : Int) : Int = {
    if(i < 0) -i else i
  } ensuring(_ $\SGE$ 0)

  def insert(in1: List, v: Int) = choose {
    (out : List) $\RA$
      content(out) $\EQ$ content(in1) $\PP$ Set(v)
  }

  def delete(in1: List, v: Int) = choose {
    (out : List) $\RA$
      content(out) $\EQ$ content(in1) $\MM$ Set(v)
  }

  def union(in1: List, in2: List) = choose {
    (out : List) $\RA$
      content(out) $\EQ$ content(in1) $\PP$ content(in2)
  }

  def diff(in1: List, in2: List) = choose {
    (out : List) $\RA$
      content(out) $\EQ$ content(in1) $\MM$ content(in2)
  }

  def split(list : List) : (List,List) = {
    choose { (res : (List,List)) $\RA$
        val s1 = size(res._1)
        val s2 = size(res._2)
        abs(s1 - s2) $\SLE$ 1 $\SAND$ s1 + s2 $\EQ$ size(list) $\SAND$
        content(res._1) $\PP$ content(res._2) $\EQ$ content(list)
     }
  }
}
\end{lstlisting}

\newpage

\subsection{SortedList}
\begin{lstlisting}
object SortedListBenchmark {
  sealed abstract class List
  case class Cons(head: Int, tail: List) extends List
  case object Nil extends List

  def size(l: List) : Int = (l match {
      case Nil $\RA$ 0
      case Cons(_, t) $\RA$ 1 + size(t)
  }) ensuring(res $\RA$ res $\SGE$ 0)

  def content(l: List): Set[Int] = l match {
    case Nil $\RA$ Set.empty[Int]
    case Cons(i, t) $\RA$ Set(i) $\PP$ content(t)
  }

  def isSorted(list : List) : Boolean = list match {
    case Nil $\RA$ true
    case Cons(_, Nil) $\RA$ true
    case Cons(x1, Cons(x2, _)) if(x1 > x2) $\RA$ false
    case Cons(_, xs) $\RA$ isSorted(xs)
  }

  def insert(in1: List, v: Int) = choose {
    (out : List) $\RA$
      isSorted(in1) $\SAND$
      (content(out) $\EQ$ content(in1) $\PP$ Set(v)) $\SAND$
      isSorted(out)
  }

  def insertAlways(in1: List, v: Int) = choose {
    (out : List) $\RA$
      isSorted(in1) $\SAND$
      (content(out) $\EQ$ content(in1) $\PP$ Set(v)) $\SAND$
      isSorted(out) $\SAND$
      size(out) $\EQ$ size(in1) + 1
  }

  def delete(in1: List, v: Int) = choose {
    (out : List) $\RA$
      isSorted(in1) $\SAND$
      (content(out) $\EQ$ content(in1) $\MM$ Set(v)) $\SAND$
      isSorted(out)
  }

  def union(in1: List, in2: List) = choose {
    (out : List) $\RA$
      isSorted(in1) $\SAND$
      isSorted(in2) $\SAND$
      (content(out) $\EQ$ content(in1) $\PP$ content(in2)) $\SAND$
      isSorted(out)
  }

  def diff(in1: List, in2: List) = choose {
    (out : List) $\RA$
      isSorted(in1) $\SAND$
      isSorted(in2) $\SAND$
      (content(out) $\EQ$ content(in1) $\MM$ content(in2)) $\SAND$
      isSorted(out)
  }

  // In order to synthesize insertionSort, we let
  // insert in the scope. Similarly for mergeSort,
  // we keep only split and union in the scope.
  def sort(list: List): List = choose {
    (res: List) $\RA$
        isSorted(res) $\SAND$
        content(res) $\EQ$ content(list)
  }

}
\end{lstlisting}

\subsection{StrictlySortedList}
\begin{lstlisting}
object Complete {
  sealed abstract class List
  case class Cons(head: Int, tail: List) extends List
  case object Nil extends List

  def size(l: List) : Int = (l match {
      case Nil $\RA$ 0
      case Cons(_, t) $\RA$ 1 + size(t)
  }) ensuring(res $\RA$ res $\SGE$ 0)

  def content(l: List): Set[Int] = l match {
    case Nil $\RA$ Set.empty[Int]
    case Cons(i, t) $\RA$ Set(i) $\PP$ content(t)
  }

  def isSorted(list : List) : Boolean = list match {
    case Nil $\RA$ true
    case Cons(_, Nil) $\RA$ true
    case Cons(x1, Cons(x2, _)) if(x1 $\SGE$ x2) $\RA$ false
    case Cons(_, xs) $\RA$ isSorted(xs)
  }

  def insert(in1: List, v: Int) = choose {
    (out : List) $\RA$
      isSorted(in1) $\SAND$
      (content(out) $\EQ$ content(in1) $\PP$ Set(v)) $\SAND$
      isSorted(out)
  }

  def delete(in1: List, v: Int) = choose {
    (out : List) $\RA$
      isSorted(in1) $\SAND$
      (content(out) $\EQ$ content(in1) $\MM$ Set(v)) $\SAND$
      isSorted(out)
  }

  def union(in1: List, in2: List) = choose {
    (out : List) $\RA$
      isSorted(in1) $\SAND$
      isSorted(in2) $\SAND$
      (content(out) $\EQ$ content(in1) $\PP$ content(in2)) $\SAND$
      isSorted(out)
  }
}
\end{lstlisting}

\newpage

\subsection{UnaryNumerals}
\begin{lstlisting}
object UnaryNumeralsBenchmark {
  sealed abstract class Num
  case object Z extends Num
  case class  S(pred: Num) extends Num

  def value(n:Num) : Int = {
    n match {
      case Z $\RA$ 0
      case S(p) $\RA$ 1 + value(p)
    }
  } ensuring (_ $\SGE$ 0)

  def add(x: Num, y: Num): Num = {
    choose { (r : Num) $\RA$
      value(r) $\EQ$ value(x) + value(y)
    }
  }

  def distinct(x: Num, y: Num): Num = {
    choose { (r : Num) $\RA$
      value(r) != value(x) $\SAND$
      value(r) != value(y)
    }
  }

  def mult(x: Num, y: Num): Num = {
    choose { (r : Num) $\RA$
      value(r) $\EQ$ value(x) * value(y)
    }
  }
}
\end{lstlisting}

\end{document}